\documentclass[aps,prl,reprint]{revtex4-1}
\usepackage{graphicx}
\usepackage{amsmath,amssymb}

\def\be{\begin{equation}}
\def\ee{\end{equation}}
\def\ber{\begin{eqnarray}}
\def\eer{\end{eqnarray}}
\def\bern{\begin{eqnarray*}}
\def\eern{\end{eqnarray*}}

\def\rv{\mathbf{r}}

\def\kv{\mathbf{k}}

\def\qv{\mathbf{q}}

\def\0v{\mathbf{0}}
\def\1v{\mathbf{1}}
\def\2v{\mathbf{2}}
\def\3v{\mathbf{3}}

\begin{document}
\title{Time-dependent localized Hartree-Fock potential}
\author {V. U. Nazarov}
\affiliation{Research Center for Applied Sciences, Academia Sinica, Taipei 11529, Taiwan}

\date{\today}
\begin{abstract}
By minimizing the difference between the left- and the right-hand sides 
of the many-body time-dependent  Schr\"{o}dinger equation with the Slater-determinant wave-function,
we derive a non-adiabatic and self-interaction free time-dependent single-particle effective potential which is the generalization to the time-dependent case of the localized Hartree-Fock potential.
The new  potential can be efficiently used within the framework of the time-dependent density-functional theory as we demonstrate by the evaluation
of the wave-vector and  frequency dependent exchange kernel $f_x(q,\omega)$ and the exchange  shear modulus $\mu_x$ of the homogeneous electron gas.
\end{abstract}
\maketitle

Time-dependent (TD) density-functional theory (DFT) \cite{Zangwill-80,*Runge-84,*Gross-85} is
in the perpetual search of effective single-particle potentials which  accurately (in ideal - exactly)
map the propagation of an interacting many-body system onto that of the non-interacting one. 
While the exchange-correlation (xc) functionals based on the Local-Density Approximation 
(LDA) \cite{Kohn-65,Perdew-92} and its semi-local refinement of the Generalized Gradient Approximation (GGA) \cite{Perdew-86,Perdew-96-0,Perdew-96,*Perdew-96-e} have proven very successful in the ground-state DFT, in TDDFT the usefulness of (semi-) local
approaches is limited due to the fundamental spatial non-locality of the exact time-dependent
xc functional \cite{Vignale-95}. 

Beyond LDA and GGA, the concept of the optimized effective potential (OEP) \cite{Sharp-53,Talman-76} plays one of the key roles in the systematic non-heuristic construction of  DFT. In the ground-state case, OEP is defined as a single-particle potential which minimizes the many-body Hamiltonian expectation value on the Slater-determinant wave-function. From the DFT perspective, OEP is the first term in the adiabatic connection series in the powers of the interaction constant (exact exchange) \cite{Gorling-94}. The generalization of the OEP to the time-dependent case has been proposed \cite{Ullrich-95}, which, however, is impractical for applications due to the formidable complexity of the integral equation involved.

In this work we propose and implement an alternative approach to the development of the time-dependent single-particle effective potential for  many-body problems. This is based on the variational principle of the minimization of the difference between the left- and right-hand sides of the time-dependent Schr\"{o}dinger equation, which principle we had introduced almost three decades ago \cite{Nazarov-85}.
By this and with no further approximations or {\it ad hoc} assumptions, we derive a time-dependent effective potential with the following useful properties: (i) It satisfies the exact large-separation asymptotic condition and is self-interaction free; (ii) It is expressed in terms of an equation easily solvable for both a finite and an infinite periodic (or homogeneous) problems;
(iii) In the static case, our effective potential reduces to the earlier known localized Hartree-Fock (HF) potential \cite{Sala-01}. As an immediate application of this approach, we derive the dynamic exchange kernel $f_x(q,\omega)$ and the exchange shear modulus $\mu_x$ \cite{Qian-02} of the homogeneous electron gas (HEG).

Let us consider a $N$-electrons system with the Hamiltonian
\begin{equation}
\hat{H}(t)= \sum\limits_i \left[-\frac{1}{2} \Delta_i +v_{ext}(\rv_i,t) \right]
+ \sum\limits_{i<j} \frac{1}{|\rv_i-\rv_j|}.
\end{equation}
The many-body wave-function $\Psi(t)$ satisfies the Schr\"{o}dinger equation
\begin{equation*}
 i \dot{\Psi} (t) = \hat{H}(t) \Psi(t), 
\end{equation*}
where the dot denotes the time derivative.
We are asking the question: What is the potential $v_{eff}(\rv,t)$ such that the functional
\begin{equation}
\int \left| i \dot{\Psi}_s(t) -\hat{H}(t) \Psi_s(t) \right|^2 d\rv_1 ... d\rv_N
\label{func}
\end{equation}
is minimal at an arbitrary time $t$ for the wave-function $\Psi_s(t)$ being the Slater determinant built with the single-particle orbitals $\psi_\alpha(\rv,t)$ which
satisfy the single-particle Schr\"{o}dinger equation
\begin{equation}
 i \dot{\psi}_\alpha(\rv,t) = \left[ -\frac{1}{2} \Delta +v_{eff}(\rv,t) \right] \psi_\alpha(\rv,t).
 \label{sp}
\end{equation}
Let us assume that the TD part of $v_{ext}(\rv,t)$ was absent at $t<0$, and we have already
solved the static problem of determining $v_{eff}(\rv)$ at $t<0$ and the corresponding orbitals 
$\psi_\alpha(\rv,t)$ at $t\le 0$. At $t=0$, the time-dependence of the potential is switched on.
Knowing the orbitals $\psi_\alpha(\rv,0)$, we find
$v_{eff}(\rv,0)$ which, determining  $\dot{\psi}_\alpha(\rv,0)$ by Eq.~(\ref{sp}), minimizes the functional (\ref{func}) at $t=0$. We then find 
$\psi_\alpha(\rv,\Delta t) \approx \psi_\alpha(\rv,0)+\dot{\psi}_\alpha(\rv,0) \Delta t$, where $\Delta t$ is a small time increase. The procedure is repeated up to an arbitrary time $t$.

In the $\Delta t\rightarrow 0$ limit, this scheme reads: With fixed (but yet unknown) orbitals $\psi_\alpha(\rv,t)$, we are looking for the potential $v_{eff}(\rv,t)$ which, determining  $\dot{\psi}_\alpha(\rv,t)$ by Eq.~(\ref{sp}), minimizes the functional (\ref{func}).
This gives $v_{eff}(\rv,t)$ as a functional of the orbitals, and, finally, the orbitals themselves are found by the self-consistent solution of Eqs.~(\ref{sp}). 
We note that a procedure of the minimization of the same functional (\ref{func})
with respect to $\dot{\psi}_\alpha$ as independently varied functions retrieves the TD HF equations  \cite{Nazarov-85}.

Since $\Psi_s(t)$  obviously satisfies the equation
\begin{equation}
i \dot{\Psi}_s(t) =  \sum\limits_i \left[-\frac{1}{2} \Delta_i +v_{eff}(\rv_i,t) \right]\Psi_s(t),
\end{equation}
the functional (\ref{func}) can be rewritten as
\begin{equation}
\int \left[ \sum\limits_i \tilde{v}(\rv_i,t)
- \sum\limits_{i<j} \frac{1}{|\rv_i-\rv_j|} \right]^2 
\left|\Psi_s(t)\right|^2
d\rv_1 ... d\rv_N,
\label{func2}
\end{equation}
where $\tilde{v}=v_{eff}-v_{ext}$. Equating to zero the first variation of Eq.~(\ref{func2})
with respect to $\delta \tilde{v}=\delta v$, we find
\begin{equation}
\begin{split}
\int \left[ \sum\limits_i \tilde{v}(\rv_i,t)
- \sum\limits_{i<j} \frac{1}{|\rv_i-\rv_j|} \right] 
\left|\Psi_s(t)\right|^2 \\ \times \sum\limits_i \delta \tilde{v}(\rv_i,t)
d\rv_1 ... d\rv_N=0,
\end{split}
\end{equation}
which can be rewritten using the permutational symmetry of the wave-function as 
\begin{equation*}
\! \int \! \left[ \sum\limits_i \tilde{v}(\rv_i,t)
\! - \! \sum\limits_{i<j} \frac{1}{|\rv_i \! - \! \rv_j|} \right] 
\left|\Psi_s(t)\right|^2  \delta \tilde{v}(\rv_1,t)
d\rv_1 ... d\rv_N \! = \! 0,
\end{equation*}
and, due to the arbitrariness of $\delta \tilde{v}$,
\begin{equation}
\int  \left[ \sum\limits_i \tilde{v}(\rv_i,t)
 -  \sum\limits_{i<j} \frac{1}{|\rv_i  - \rv_j|} \right]  
\left|\Psi_s(t)\right|^2  
d\rv_2 ... d\rv_N  =    0.
\label{e7}
\end{equation}
Straightforward but rather lengthy  transformations carried out in  Ref.~\footnote{See EPAPS Document No  ...}\newcounter{ft}\setcounter{ft}{\value{footnote}}
lead  from Eq.~(\ref{e7}) to the following equation for the
exchange potential $v_x=\tilde{v}-v_H$, where $v_H(\rv)$ is the Hartree potential,
\begin{equation}
\begin{split}
n(\rv,t) v_x(\rv,t)=
 \int  \left[ v_x(\rv_1,t)
\! - \!  \frac{1}{|\rv-\rv_1|} \right] 
|\rho(\rv,\rv_1,t)|^2
d\rv_1  \\
 +  \int    
 \frac{\rho(\rv,\rv_1,t)\rho(\rv_1,\rv_2,t)\rho(\rv_2,\rv,t)}{|\rv_1-\rv_2|}  
d\rv_1  d\rv_2,
\end{split}
\label{main4}
\end{equation}
where
\begin{align}
&n(\rv,t)=\sum\limits_{\alpha=1}^N |\psi_\alpha(\rv,t)|^2,\\
&\rho(\rv,\rv_1,t)=\sum\limits_{\alpha=1}^{N} \psi_\alpha(\rv,t) \psi_\alpha^*(\rv_1,t) 
\label{dm}
\end{align}
are the particle density and the single-particle density-matrix, respectively
\footnote{In Eq.~(\ref{main4}), the integration over the space coordinates also implies
the summation over spin indices.}.
Equation (\ref{main4}) is our main result. We note that the only difference
of Eq.~(\ref{main4}) from the earlier known equation for the {\em static} 
localized HF potential (see Refs. \onlinecite{Sala-01} and \onlinecite{Zhou-05} for spin-neutral and spin-polarized cases, respectively) 
is the time-dependence of all the quantities involved.
It must, however, be emphasized that without the derivation from
the time-dependent variational principle, the generalization of the static
localized HF potential to the time-dependent case by just inserting the time variable into the static equation would have been ungrounded.
Trivially, Eq.~(\ref{main4}) reduces to the equation for the localized HF potential
in the time-independent case.

Solution of Eq.~(\ref{main4}) in the case of a few-body system does not present a difficult problem,
as have already been pointed out in Ref.~\onlinecite{Sala-01} in conjunction with the time-independent
case: Due to Eq.~(\ref{dm}), the kernel $|\rho(\rv,\rv_1,t)|^2$ of the integral equation (\ref{main4}) is a separable function with respect to $\rv$ and $\rv'$ variables. Neither presents
it a problem in the case of infinite periodic systems, when the equation reduces to the matrix one.
We now  apply the TD localized HF potential to obtain the wave-vector
and frequency dependent exchange kernel $f_x(q,\omega)$ of HEG, which, being a fundamental
quantity by itself, is also an important input in the theory of optical response
of a weakly ingomogeneous interacting electron gas \cite{Nazarov-09}.

{\em Dynamic exchange kernel of HEG~--}
In the case of HEG and a weak externally applied potential $\delta v_{ext}(\rv,t)= \delta v_{ext}(\qv,\omega) e^{i(\qv\cdot \rv-\omega t)}$, we linearize Eq.~(\ref{main4}) 
with respect to the latter potential. The zero-order orbitals are plane-waves
\footnote{Caution must be exercised when solving the ground-state problem for HEG with Eq.~(\ref{main4}): Although the ground-state
effective potential is constant, it occurs to be infinite for the infinite system. A proper limiting procedure, however,
of starting from a finite volume with periodic boundary conditions, resolves this difficulty unambiguously. }, 
and to the zeroth and first orders we have for the density-matrix
\begin{equation}
\begin{split}
&\rho_0(\rv,\rv_1)= \frac{2}{V} \sum\limits_{\kv}  \int f(\epsilon_\kv) e^{i \kv\cdot(\rv-\rv_1)} ,\\
&\delta\rho(\rv,\rv_1,\omega) \! = \! \delta v_s(\qv,\omega) \, e^{i \qv \cdot\rv} 
\frac{2}{V} \! \sum\limits_{\kv} \! \frac{f(\epsilon_\kv) \! - \! f(\epsilon_{\kv+\qv})}{\omega \!  - \!  \epsilon_{\kv+\qv}  \! + \!  \epsilon_{\kv}  \! + \! i\eta} e^{i\kv\cdot(\rv \! - \! \rv_1)},
\end{split}
\end{equation}
where $\epsilon_\kv=k^2/2$ are free-particle eigenenergies, $f(\epsilon_{\kv})$ are  their occupation numbers, 
$\delta v_s(\qv,\omega)$ is the perturbation of the Kohn-Sham (KS) potential,
$V$ is the normalization volume, and $\eta$ is an infinitesimal positive. After linearization, Eq.~(\ref{main4}) yields

\begin{widetext}
\begin{equation}
\begin{split}
\frac{ \delta v_x(q,\omega)}{\delta v_s(q,\omega)}=  \frac{4\pi}{(2\pi)^3}
\int 
\left[ \frac{1}{\omega \! - \! \epsilon_{\kv+\qv} \! + \! \epsilon_{\kv} \! + \! i\eta} -
\frac{1}{\omega \! + \! \epsilon_{\kv+\qv} \! - \! \epsilon_{\kv} \! + \! i\eta} \right] 
\frac{  f(\epsilon_\kv) B(\qv,\kv) -f(\epsilon_{\kv+\qv}) [f(\epsilon_\kv)-1] C(k)
 }{n_0- A(q) } d\kv,
\end{split}
\label{HEG}
\end{equation}
\end{widetext}
where
\begin{align}
&A(q)=\frac{2}{(2\pi)^3} \int f(\epsilon_\kv) f(\epsilon_{\kv+\qv}) d\kv,
\label{AI} \\
&B(\qv,\kv)=\frac{2}{(2\pi)^3} \int \frac{f(\epsilon_{\kv_1})f(\epsilon_{\kv_1+\qv})}{|\kv-\kv_1|^2} d\kv_1,
\label{BI} \\
&C(k)= \frac{2}{(2\pi)^3} \int \frac{f(\epsilon_{\kv_1})}{|\kv-\kv_1|^2} d\kv_1.
\label{CI}
\end{align}
With the use of Eq.~(\ref{HEG}), the exchange kernel is now found as
\begin{equation}
\begin{split}
&f^h_x(q,\omega) \equiv \frac{\delta v_x(q,\omega)}{\delta n(q,\omega)}=\\
&\frac{\delta v_x(q,\omega)}{\delta v_s(q,\omega)} \frac{\delta v_s(q,\omega)}{\delta n(q,\omega)}=
\frac{\delta v_x(q,\omega)}{\delta v_s(q,\omega)}  \left(\chi_s^h\right)^{-1}(q,\omega),
\end{split}
\label{fv}
\end{equation}
where $\chi^h_s(q,\omega)$ is Lindhard density-response function \cite{Lindhard-54}
\begin{equation}
\chi^h_s(q,\omega)=\frac{2}{V} \sum\limits_{\kv} \frac{f(\epsilon_\kv) \! - \! f(\epsilon_{\kv+\qv})}{\omega \!  - \!  \epsilon_{\kv+\qv}  \! + \!  \epsilon_{\kv}  \! + \! i\eta} .
\label{chis}
\end{equation}
In Ref.~\footnotemark[\value{ft}], we evaluate integrals (\ref{AI}) and (\ref{CI}) analytically
and reduce the integral (\ref{BI}) to a single-fold one.

It can be seen from Eqs.~(\ref{HEG}) and (\ref{fv})  that ${\rm Im} \, f_x$ is nonzero inside 
the single particle-hole excitation continuum only. This deficiency of $f_x$ derived from TD
localized HF potential is not limited to HEG, but, as can be easily seen, persists
for any extended (periodic) system. Therefore, such subtle effect as the high-frequency tail
of ${\rm Im} \, f_x$ of HEG \cite{Sturm-00} cannot be accounted for within the present approach.
Instead, we will now show that $f_x$ derived from the TD localized HF potential significantly corrects the Lindhard dielectric function of HEG within the single particle-hole continuum.

\begin{figure}[h] 
\includegraphics[width=  \columnwidth, trim=50 0 0 0, clip=true]{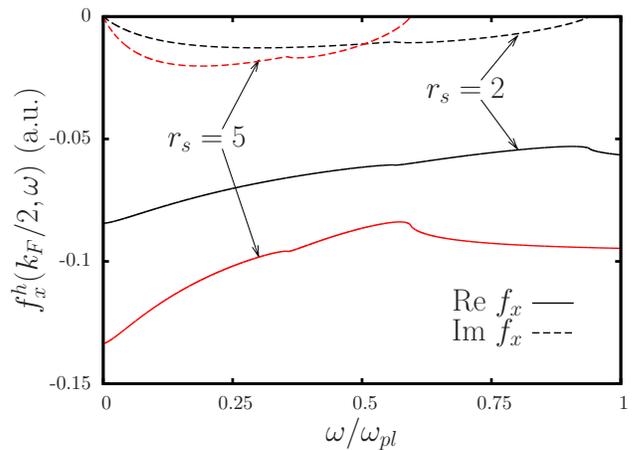} 
\caption{\label{fx_fig} (color online)
Exchange kernel $f^h_x(q,\omega)$, $q=0.5 \times k_F$, of HEG of $r_s=2$ (black curves online) and $r_s=5$ (red curves online). Solid and dashed curves are real and imaginary parts of $f^h_x$.
}
\end{figure}

In Fig.~\ref{fx_fig}, the exchange kernel $f^h_x(q,\omega)$ obtained by the use of Eqs.~(\ref{HEG}) and (\ref{fv}) is plotted at $q=0.5 \times k_F$ for $r_s=2$ and $5$.
With the inclusion of $f_x$, the dielectric function of HEG can be written as
\begin{equation}
\epsilon^h(q,\omega)=1-\frac{4\pi}{q^2} \frac{\chi^h_s(q,\omega)}{1+\chi^h_s(q,\omega) f^h_x(q,\omega)}.
\label{eps}
\end{equation}

\begin{figure}[h] 
\includegraphics[width=  \columnwidth, trim=20 0 0 0, clip=true]{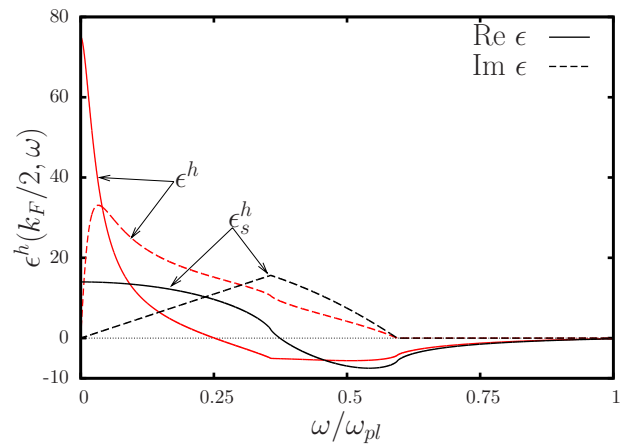} 
\caption{\label{eps_fig} (color online)
Dielectric function $\epsilon^h(q,\omega)$, $q=0.5 \times k_F$, of HEG of $r_s=5$
evaluated by the use of EQ.~(\ref{eps})
with the exchange kernel $f^h_x$ included (red curves online) and its Lindhard counterpart
$\epsilon^h_s(q,\omega)$ (black curves online). Solid and dashed curves are real and imaginary
parts of the dielectric function, respectively.
}
\end{figure}

In Fig.~\ref{eps_fig}, the dielectric function of HEG of $r_s=5$ obtained through
Eq.~(\ref{eps}) is plotted together with the Lindhard dielectric function.
From this we judge that dynamic exchange plays a significant role at this density
and  can hardly be considered as a weak perturbation.

\begin{figure}[h] 
\includegraphics[width= \columnwidth, trim=30 0 0 0, clip=true]{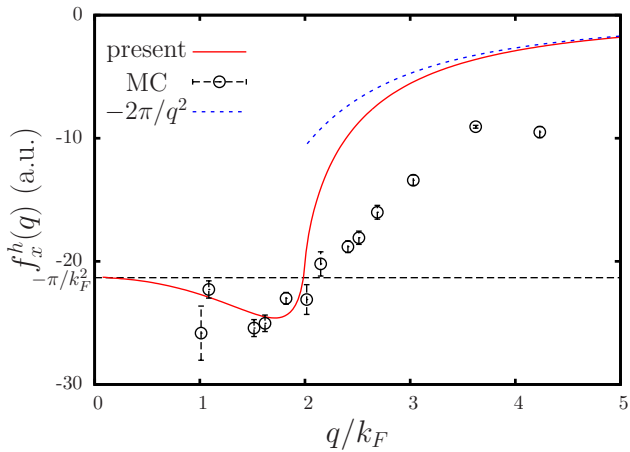} 
\caption{\label{stat} (color online)
Static exchange kernel $f^h_x(q)$ of HEG of $r_s=5$.
Solid line (red online) is the present result.
Symbols with error bars are $f^h_{xc}(q)$ by MC simulations from Ref. \cite{Moroni-95}.
The dashed line (blue online) shows the asymptotic behavior at large $q$,
as stipulated by Eq.~(\ref{statasy}).
}
\end{figure}

In Fig.~\ref{stat}, the static exchange kernel $f^h_x(q)$ is plotted for HEG of $r_s=5$
as a function of the wave-vector. We find a qualitative agreement with the  Monte Carlo (MC)
simulations of Ref. \onlinecite{Moroni-95}.

The following limiting cases can be further worked out from Eqs.~(\ref{HEG})-(\ref{CI})
\footnote{It must be noted that in Eq.~(\ref{qw}) we have not been able to evaluate the numerator
analytically, but rather, having evaluated it numerically to $3.1415$, surmised it to be $\pi$. This, however, does not affect the following discussion at all.}
\begin{align}
& \lim_{q\rightarrow 0}
f_x^h(q,\omega\ne 0)= -\frac{3\pi}{4 k_F^2},
\label{wq} \\
&\lim_{q\rightarrow 0} \lim_{\omega\rightarrow 0} 
f_x^h(q,\omega)= -\frac{\pi}{k_F^2},
\label{qw} \\
&\lim_{q\rightarrow \infty} \lim_{\omega\rightarrow 0} 
f_x^h(q,\omega)= -\frac{2 \pi}{q^2},
\label{statasy}
\end{align}
and since \cite{Qian-02}
\begin{equation}
\left[\lim_{\omega\rightarrow 0} \lim_{q\rightarrow 0} -
\lim_{q\rightarrow 0} \lim_{\omega\rightarrow 0} \right] f_x^h(q,\omega) =
\frac{4}{3}  \frac{\mu_{xc}}{n^2 },
\label{mu0}
\end{equation}
where $\mu_x$ is the exchange contribution to
the shear modulus of HEG, we can write by virtue of Eqs.~(\ref{wq}), (\ref{qw}), and (\ref{mu0})
\begin{equation}
\mu_x=\frac{3 \left(\frac{3}{2}\right)^{2/3}}{64 \pi ^{5/3} r_s^4}\approx \frac{0.0091}{r_s^4},
\label{mux}
\end{equation}
which is smaller than the high-density result of Ref.~\onlinecite{Conti-99}: 
$\mu_{xc}=\frac{3 \left(\frac{3}{2}\right)^{2/3}}{40 \pi ^{5/3} r_s^4}\approx \frac{0,0146}{r_s^4}$.
In Table~\ref{tab_mx}, we compare $\mu_x$ obtained via Eq.~(\ref{mux})
with $\mu_{xc}$ of Refs.~\onlinecite{Qian-02}, \onlinecite{Bohm-96,*Conti-97,*Nifosi-97,*Nifosi-98},
and \onlinecite{Conti-99}.
\begin{table}[h]
\caption{\label{tab_mx} Exchange shear modulus $\mu_x$ in units of $2 \omega_{pl} n$.
Present results are shown together with those for $\mu_{xc}$ of Refs. 
\onlinecite{Qian-02} (QV), \onlinecite{Bohm-96,*Conti-97,*Nifosi-97,*Nifosi-98} (NCT), and \onlinecite{Conti-99} (CV).}
\begin{tabular}{llllll}
\hline\hline
$r_s$ & 1 & 2 & 3 & 4 & 5 \\
\hline
present & 0.01102 & 0.01559 & 0.01909 & 0.02204 & 0.02464 \\
QV & 0.00738 & 0.00770 & 0.00801 & 0.00837 & 0.00851 \\
NCT & 0.0064 & 0.052 & 0.0037 & 0.0020 & 0.0002 \\
CV & 0.01763 & 0.02494 & 0.03054 & 0.03527 & 0.03943 \\
\hline\hline
\end{tabular}
\end{table}

In conclusion, within the well-defined procedure
of the minimization of the difference between the left- and right-hand sides of the time-dependent
Shr\"{o}dinger equation, we have derived a time-dependent single-particle effective potential
for a system of arbitrary number of electrons under the action of a time-dependent external field.
This potential is non-local, non-adiabatic, and self-interaction free, and it satisfies the exact large-separation asymptotic condition. 
At the same time, our effective potential is comparatively easy for evaluation,
which is in contrast to earlier known TD optimized effective potential.
These properties open a way to efficiently use this potential within the context of time-dependent density-functional theory, 
as we demonstrate by the derivation of the exchange kernel $f_x(q,\omega)$ of the homogeneous electron gas.

\acknowledgements
I acknowledge support from National Science Council, Taiwan, Grant No. 100-2112-M-001-025-MY3.

%

\onecolumngrid

\newpage

\thispagestyle{empty}

\
\renewcommand{\theequation}{{S.\arabic{equation}}}
\renewcommand{\thefigure}{{S.\arabic{figure}}}

\section{SUPPLEMENTAL MATERIAL\\
\ \\
to the paper by V. U. Nazarov \\
\ \\
``Time-dependent localized Hartree-Fock potential''
}

\setcounter{page}{1}
\setcounter{equation}{0}
\setcounter{figure}{0}

\subsection{Derivation of Eq.~(\ref{main4}).}

Using the permutational symmetry, we can rewrite Eq. (\ref{e7}) as
\begin{equation}
\tilde{v}(\rv_1,t) \int \left|\Psi_s(t)\right|^2  
d\rv_2 ... d\rv_N  
+ \int \left[ (N-1) \tilde{v}(\rv_2,t)
-  \frac{(N-1)}{|\rv_1-\rv_2|}  
-  \frac{(N-1)(N-2)}{2 |\rv_2-\rv_3|} \right] 
\left|\Psi_s(t)\right|^2  
d\rv_2 ... d\rv_N=0,
\end{equation}
or
\begin{equation}
\frac{\tilde{v}(\rv_1,t) n(\rv_1,t)}{N}
+ (N-1) \int \left[ \tilde{v}(\rv_2,t)
-  \frac{1}{|\rv_1-\rv_2|} 
-  \frac{N-2}{2 |\rv_2-\rv_3|} \right] 
\left|\Psi_s(t)\right|^2  
d\rv_2 ... d\rv_N=0,
\end{equation}
where $n(\rv,t)$ is the particle-density. Further simplifications give
\begin{equation}
 \frac{\tilde{v}(\rv_1,t)}{N(N-1)} 
\! + \!   \int \! \left[ \tilde{v}(\rv_2,t)
\! - \!  \frac{1}{|\rv_1-\rv_2|} \right] 
  \frac{\rho_2(\rv_1,\rv_2;\rv_1,\rv_2;t)}{n(\rv_1,t)}
d\rv_2   
- \frac{(N \! - \! 2)}{2} \! \int    
 \! \frac{\rho_3(\rv_1,\rv_2,\rv_3;\rv_1,\rv_2,\rv_3;t)}{n(\rv_1,t) |\rv_2-\rv_3|}  
d\rv_2  d\rv_3 \! = \! 0,
\label{main}
\end{equation}
where the $k$-particle density-matrix is
\begin{equation}
\rho_k(\rv_1,...\rv_k;\rv_1',...\rv_k';t)= 
\int \Psi_s(\rv_1...\rv_k,\rv_{k+1}''...\rv_N'',t)
\Psi_s^*(\rv_1'...\rv_k',\rv_{k+1}''...\rv_N'',t) d\rv_{k+1}''...d\rv_N''.
\end{equation}
For the Slater determinant wave-function $\Psi_s(t)$ the equalities hold
\begin{align}
&\rho_2(\rv_1,\rv_2;\rv_1,\rv_2;t)  = \frac{1}{N(N \!-1 \!)} 
\left[ n(\rv_1,t) n(\rv_2,t)
 -|\rho(\rv_1;\rv_2,t)|^2 \right], \\
\begin{split}
& \rho_3(\rv_1,\rv_2,\rv_3;\rv_1,\rv_2,\rv_3;t)  = \frac{1}{N(N \!-1 \!)(N \!- \! 2)} 
 \left[ n(\rv_1,t) n(\rv_2,t) n(\rv_3,t)  
-n(\rv_2,t) |\rho(\rv_1;\rv_3,t)|^2 \right. 
-n(\rv_3,t) |\rho(\rv_1;\rv_2,t)|^2 \\
&-n(\rv_1,t) |\rho(\rv_2;\rv_3,t)|^2 +\rho(\rv_1;\rv_2,t)\rho(\rv_2;\rv_3,t)\rho(\rv_3;\rv_1,t)
\left. +\rho(\rv_1;\rv_3,t)\rho(\rv_3;\rv_2,t)\rho(\rv_2;\rv_1,t) \right],
\end{split}
\end{align}
where 
\begin{equation}
\rho(\rv;\rv',t)=N \rho_1(\rv;\rv',t)=\sum\limits_\alpha \psi_\alpha(\rv,t) \psi_\alpha^*(\rv',t) . 
\end{equation}
Therefore
\begin{equation}
\begin{split}
\tilde{v}(\rv_1,t)
-V_H(\rv_1,t) 
\! - \!   \int \! \left[ \tilde{v}(\rv_2,t)
\! - \!  \frac{1}{|\rv_1-\rv_2|} \right] 
\frac{|\rho(\rv_1;\rv_2;t)|^2}{n(\rv_1,t)}
d\rv_2  
+ \! \int    
 \! \frac{n(\rv_2,t) |\rho(\rv_1;\rv_3,t)|^2}{n(\rv_1,t) |\rv_2-\rv_3|}  
d\rv_2  d\rv_3
 \\
-  \! \int    
 \! \frac{\rho(\rv_1;\rv_2,t)\rho(\rv_2;\rv_3,t)\rho(\rv_3;\rv_1,t)}{n(\rv_1,t) |\rv_2-\rv_3|}  
d\rv_2  d\rv_3 +C(t) \! = \! 0, \
\end{split}
\label{main2}
\end{equation}
where
\begin{align}
&V_H(\rv,t)=\int \frac{n(\rv',t)}{|\rv-\rv'|} d\rv',\\
&C(t)=\int \tilde{v}(\rv_2,t) n(\rv_2,t) d\rv_2 
- \frac{1}{2} \! \int    
 \! \frac{n(\rv_2,t) n(\rv_3,t)}{ |\rv_2-\rv_3|}  
d\rv_2  d\rv_3 +\frac{1}{2} \! \int    
 \! \frac{|\rho(\rv_2;\rv_3,t)|^2}{ |\rv_2-\rv_3|}  
d\rv_2  d\rv_3=0,
\label{wC}
\end{align}
and the second equality in Eq.~(\ref{wC}) follows from Eq.~(\ref{main2}). Eqs.~(\ref{main2}) and (\ref{wC}) 
give immediately  Eq.~(\ref{main4}).

\subsection{Integrals (\ref{AI})-(\ref{CI}).}

For integrals (\ref{AI}) and (\ref{CI}) we have straightforwardly
\begin{align}
&A(q) = 
\frac{1}{48 \pi^2} \Theta(2 k_F-q) (2 k_F-q)^2 (4 k_F+q),\\
&C(k)=\frac{2}{(2\pi)^3} H(k,k_F),
\end{align}
where $\Theta(x)$ is the Heaviside's step function and
\begin{equation}
H(k,p)=\frac{\pi}{ k} \left[
(p^2-k^2) \log \left|\frac{p+k}{p-k}\right|+2 k p \right].
\end{equation}

For integral (\ref{BI}) we have

\begin{equation}
B(\qv,\kv)=  
\frac{2}{(2\pi)^3} \Theta(k_F-q) H(k,k_F-q) 
+\frac{2}{(2\pi)^2} \Theta(2 k_F-q) \int\limits_{|k_F-q|}^{k_F}   
P\left(k,k_1,\frac{k_F^2-k_1^2-q^2}{2 k_1 q},\frac{\kv\cdot\qv}{k q}\right) d k_1,
\end{equation}
where
\begin{equation}
\begin{split}
P(k,k_1,x,y) & =  \frac{k_1} {2 k} \left\{\log \left[ \sqrt{k^4+4 k^2 k_1^2
   x^2+2 k k_1 \left(k k_1 \left(2 y^2-1\right)-2 x y
   \left(k^2+k_1^2\right)\right)+k_1^4}+2 k k_1 x-y \left(k^2+k_1^2\right)\right] \right. \\ & \left. -  \log \left(
   (1-y) (k-k_1)^2\right)\right\}
   \end{split}
\end{equation}

\end{document}